\documentclass[prb,twocolumn,showpacs]{revtex4}
\usepackage{amsmath}
\usepackage{dcolumn}
\usepackage{graphicx}

\begin{document}

\title[Short title for running header]{Pairing origin of the pseudogap as observed in ARPES measurement in the underdoped cuprates}
\author{Tao Li and Da-Wei Yao}
\affiliation{Department of Physics, Renmin
University of China, Beijing 100872, P.R.China}
\date{\today}

\begin{abstract}
We show that electron pairing is indispensable for the development of the leading edge gap as observed in ARPES measurement in the underdoped cuprates, even though clear evidence for the violation of the particle-hole symmetry is found in the electron spectrum. To support this assertion, we studied the electron spectrum under the scattering of diffusive antiferromagnetic(AF) spin fluctuation, which is thought to be a major candidate for a competing order in the competing order scenario of the pseudogap phenomena. We find that the Fermi level crossing along the M=$(\pi,0)$ to X=$(\pi,\pi)$ line can only be avoided when the M point is pushed above the Fermi level in this scenario. We argue that the same conclusion holds in all competing order scenarios that  preserve the U(1) charge conservation. The inconsistency between this prediction and the ARPES observation implies that a competing order in the particle-hole channel alone is not sufficient to explain the pseudogap as observed in ARPES measurement. We also find that the electron system always forms a single large Fermi surface under the scattering of short-ranged dynamical spin fluctuation, rather than forming small Fermi pockets as predicted by the AF band folding picture. The AF shadow band is smeared out in energy as a result of the dispersion in the scattered quasiparticle state and the diffusion in spin fluctuation energy. Nevertheless, we find that the AF band folding effect is important for the understanding of the quasiparticle dynamics in the pseudogap phase, especially, of the origin of the high energy hump structure in the anti-nodal region and the signature of particle-hole asymmetry in the electron spectrum. It may even provide the driving force of the pseudogap phenomena, since the strong dressing of the anti-nodal quasiparticle by the AF spin fluctuation will greatly reduce the kinetic energy penalty for electron pairing in this region. 
\end{abstract}


\maketitle

One of the main unresolved issue in the study of high T$_{c}$ cuprates is the origin of the pseudogap phenomena, which is  observed ubiquitously in the quasiparticle spectrum, the spin fluctuation spectrum, transport properties and thermodynamic properties of the system\cite{Timusk,ARPES1,Lee,Kordyuk}. There are two types of scenario for the origin of the pseudogap phenomena. The first type of scenario takes the pseudogap phenomena as a superconducting fluctuation effect. While supported by several measurements\cite{Orenstein,Xu,Lu}, the superconducting fluctuation scenario is generally believed to be insufficient to account for the whole complexity of the pseudogap phenomena. The second type of scenario relates the pseudogap phenomena to an order parameter that is competing with the superconducting order, which usually preserves the U(1) charge conservation. 

To understand the origin of the pseudogap phenomena, it is important to know how the different manifestations of the pseudogap phenomena are related to each other. Here we would like to distinguish two types of pseudogap phenomena discussed in the literature\cite{Weak}. The so called upper pseudogap begins at a characteristic temperature T$_{0}$, below which the uniform susceptibility of the system begins to decrease with temperature. At the same time, antiferromagnetic spin fluctuation begins to develop, as evidenced by the increase of the spin relaxation rate $1/^{63}\mathrm{T}_{1}\mathrm{T}$. The so called lower pseudogap begins at a characteristic temperature T*, where the spin relaxation rate reaches a maximum.

Recently, ARPES measurement has provided with unprecedented detail the evolution of the electron spectral function with temperature in the underdoped cuprates\cite{Particle,ARPES6,ARPES7}. In particular, it exposes for the first time how the pseudogap is developed on a large complete Fermi surface around T*, at which the spin relaxation rate $1/^{63}\mathrm{T}_{1}\mathrm{T}$ reaches its maximum\cite{Zheng}.  Below T*, part of the Fermi surface in the anti-nodal region is eliminated, leaving the system with a open Fermi arc in the nodal region. The electron spectrum in the anti-nodal region is characterized by the development of a weakly momentum dependent leading edge gap and the emergence of a broad hump structure at higher energy. It is found that the spectral maximum of the hump structure bends back at a momentum $k_{\mathrm{G}}$ that is different from the Fermi momentum $k_{\mathrm{F}}$ above T*. This mismatch is interpreted as a strong evidence against the pairing origin of the pseudogap, which predicts that the minimal gap should be achieved on the underlying Fermi surface, and is regarded as being suggestive of a competing order origin of the pseudogap.

The antiferromagnetic(AF) spin fluctuation scenario is the most extensively studied competing order scenario for the origin of the pseudogap phenomena. The existence of the strong AF spin fluctuation in the cupartes has been well documented by early neutron scattering and NMR measurements. More recently, RIXS measurements found that the high energy spin fluctuation in the cuprates is hardly changed from that of the AF insulating parent compounds by carrier doping\cite{RIXS1,RIXS2,RIXS3,RIXS4,RIXS5,RIXS6}, indicating that the electron spin in the system is behaving more like a quantized local moment. The dual nature of the electrons as both itinerant quasiparticles and quantized local moments in the cuprates leaves us a formidable task to develop a theory of the high temperature superconductivity. However, at the phenomenological level, the quasiparticle and the local moment behavior of the electron can be treated as independent degree of freedoms that are coupled by a phenomenological interaction. The result of such an abstraction is called the spin-Fermion model\cite{NAFL1}.  At a microscopic level, the spin-Fermion model can be understood as the low energy effective theory of a strongly correlated electron system, in which the local moment results from the integration over the high energy Fermions. In the spin-Fermion picture, the AF spin fluctuation is not only the pairing glue of the quasiparticles, but is also thought to be responsible for the non-Fermi liquid behavior and even the pseudogap phenomena of the cuprates\cite{NAFL2,NAFL3,NAFL4,NAFL5}.

In the AF spin fluctuation scenario, the pseudogap in the electron spectrum has been proposed to be the result of the AF band folding effect, which is illustrated in Figure 1 for system with AF long range order. More specifically, the Fermi surface of the system will be reconstructed into closed Fermi pockets as a result of the level repulsion effect between the bare quasiparticle band and the AF shadow band. The Fermi pockets around the M point will shrink in size with the increase of the AF order and will disappear when the M point is pushed above the Fermi level. In such a case, the Fermi level crossing along the M-X line will be eliminated, leaving the system with a back bending point in the dispersion of the AF shadow band at a finite binding energy(see Figure 2). The back bending momentum is in general different from the Fermi momentum of the underlying Fermi surface. 

These predictions of the AF band folding picture, however, are not quite consistent with the ARPES observation. First, the Fermi momentum along the M-X line is found to be almost temperature independent above T* and the M point is found to be always below the Fermi level. Second, rather than a closed Fermi pocket, a open Fermi arc is observed in the nodal region. In other words, the AF shadow of the Fermi surface seems to be totally missing.

\begin{figure}
\includegraphics[width=10cm]{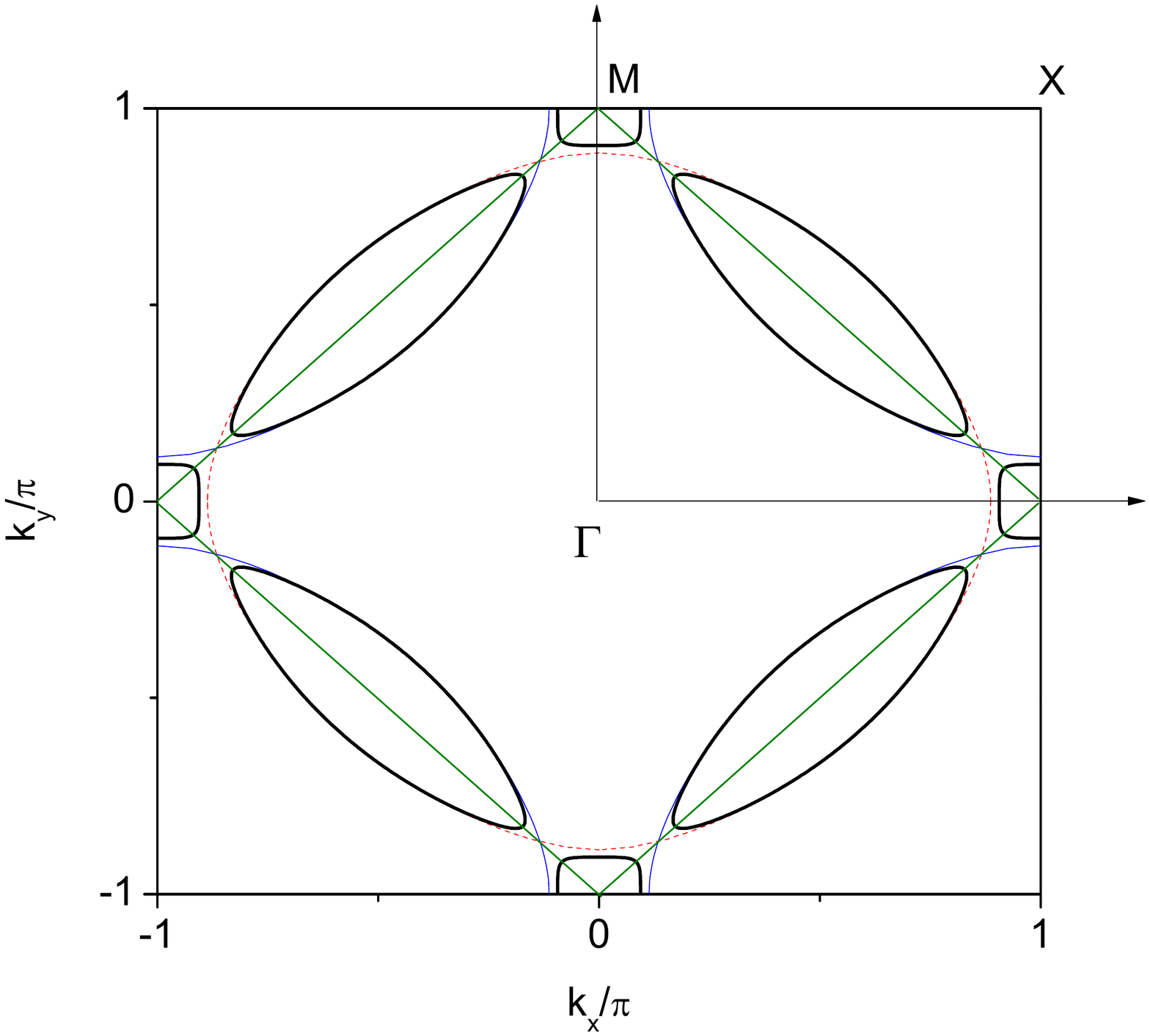}
\caption{\label{fig1}
(Color on-line) Illustration of the Fermi pockets and the hot spots in the AF band folding picture. The blue line indicates the bare Fermi surface, the red dashed line indicates its shadow under the scattering of the AF long range order. The bare Fermi surface and its AF shadow intersect at the hot spots. The black line indicates the reconstructed Fermi surface in the presence of the AF long range order. The green line marks the boundary of the AF Brillouin zone.}
\end{figure}

\begin{figure}
\includegraphics[width=8cm]{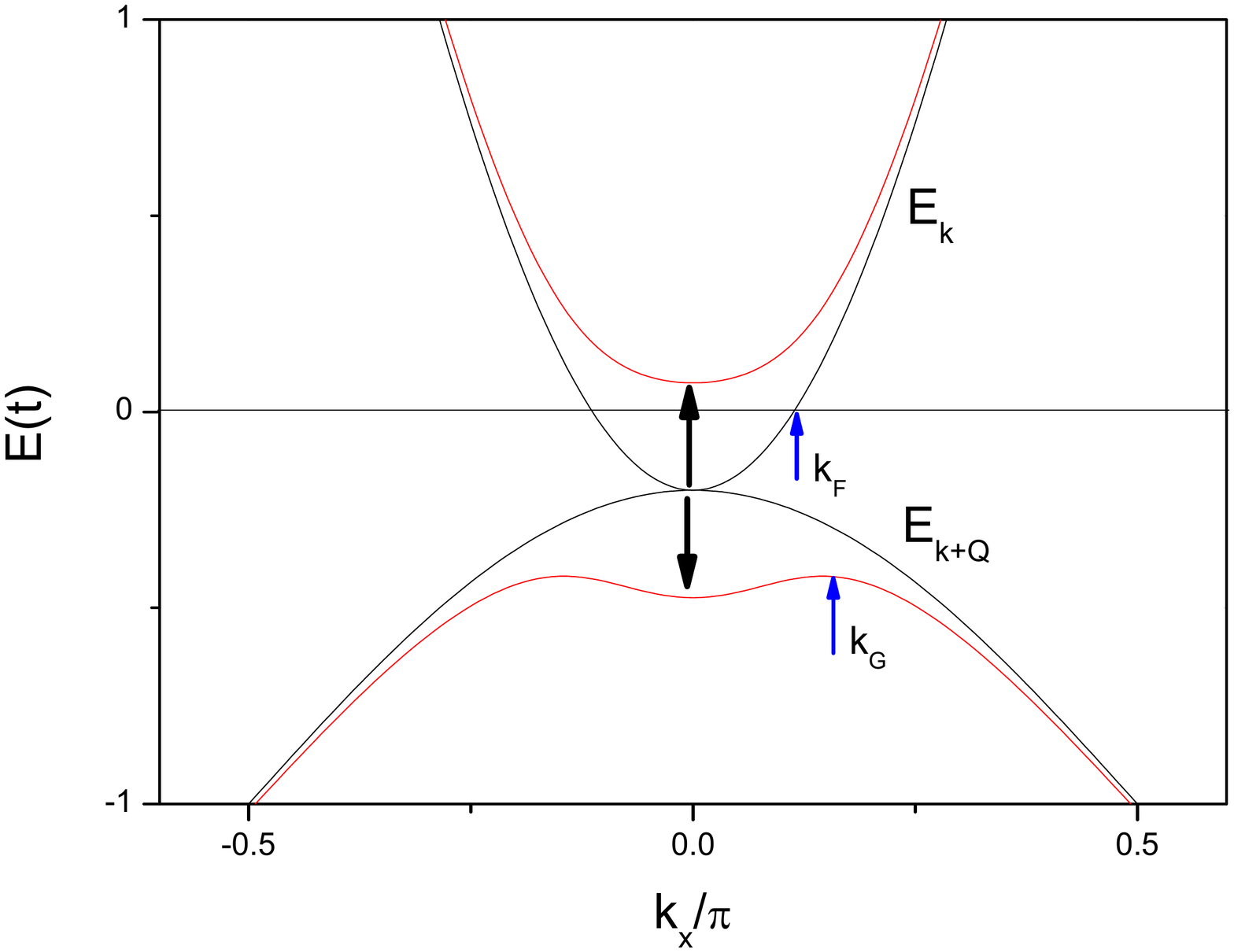}
\caption{\label{fig1}
(Color on-line) Illustration of the level repulsion effect between the bare band at $\epsilon_{\mathrm{k}}$ and the AF shadow band at $\epsilon_{\mathrm{k+Q}}$ in the anti-nodal region. The momentum is along the M-X line. The black line marks the position of the Fermi level. The blue arrows mark the Fermi momentum $k_{\mathrm{F}}$ and the back bending momentum $k_{\mathrm{G}}$. }
\end{figure}

A possible excuse for the failure of the AF band folding picture presented above is the over-simplification on the AF spin fluctuation in the cuprates, which is both short-ranged and dynamical. In the background of short-ranged and dynamical spin fluctuation, the scattered state will have an energy that is uncertain both because of the quasiparticle dispersion in $\epsilon_{\mathrm{k+q}}$ and the diffusion in the spin fluctuation energy $\Omega_{\mathrm{q}}$\cite{edop}. The AF shadow band will be obscured when such an effect is strong enough. We note that the dispersion effect is extremely anisotropic on the Fermi surface. It is weakest in the anti-nodal region as a result of the proximity to the Van Hove singularity and strongest in the nodal region. More specifically, the energy uncertainty of scattered state caused by dispersion effect is given roughly by $2\pi v_{\mathrm{F}}/\xi$ in the nodal region, where $v_{\mathrm{F}}$ is the Fermi velocity in the nodal region and $\xi$ is the spin correlation length\cite{edop}. If we assume $\xi=3$, which is typical for the underdoped cuprates, the energy uncertainty is about half of the total band width. The shadow band is certainly ill-defined in such a case. The naive AF band folding picture is thus not applicable to the hole-doped cuprate systems. This motivates us to study the quasiparticle dynamics with more realistic spin fluctuation parameters.

 The spin-Fermion model takes the form of
 \begin{equation}
 H=\sum_{\mathrm{k},\sigma}\epsilon_{\mathrm{k}}c^{\dagger}_{\mathrm{k},\sigma}c_{\mathrm{k},\sigma}+g\sum_{i}\vec{\mathrm{S}}_{i}\cdot\vec{\mathrm{s}_{i}}.\nonumber
 \end{equation}
Here $\epsilon_{\mathrm{k}}=-2t(\cos k_{x}+\cos k_{y})-4t'\cos k_{x}\cos k_{y}-\mu$ is the bare dispersion of the quasiparticle. $\vec{\mathrm{s}}_{i}=\frac{1}{2}\sum_{\alpha,\beta}c^{\dagger}_{i,\alpha}\vec{\sigma}_{\alpha,\beta}c_{i,\beta}$ is the spin density operator of the itinerant electron and $\vec{\mathrm{S}}_{i}$ is the local moment operator. $g$ is a phenomenological coupling constant. In this study, we set $t=250 \ meV$, $t'=-0.3t$ and $\mu=-t$, so that the doping level is about $x=0.14$. At the same time, we will adopt the widely used Monien-Mills-Pines(MMP) form as a phenomenological guess of the dynamical spin susceptibility of the local moment\cite{MMP1,MMP2}, which is given by
 \begin{equation}
\chi(\mathrm{q},\omega)=\frac{\chi_{\mathrm{Q}}}{1+(\mathrm{q}-\mathrm{Q})^{2}\xi^{2}-i\omega/\omega_{sf}}.\nonumber
 \end{equation}
Here $\chi_{\mathrm{Q}}\propto \xi^{2}$ is the static spin susceptibility at the antiferromagnetic wave vector $\mathrm{Q}=(\pi,\pi)$, $\xi$ is the spin correlation length, $\omega_{sf}$ is a characteristic frequency describing the dissipation of the local moment by its coupling to the itinerant quasiparticles. In this study, we set $\xi=3a $ and $\omega_{sf}= 15 \ meV$ at a temperature of $k_{\mathrm{B}}\mathrm{T}= t/20 $, which are typical for underdoped cuprates. Lastly, we note that the integrated spectral weight of the MMP susceptibility actually diverges logarithmically at high energy. To remove such an unphysical divergence, we cut off the spectral weight at $\omega_{c}=30  \omega_{sf}=450 \ meV$. Such a choice for $\omega_{c}$ is consistent with the RIXS measurement on high T$_{c}$ cuprates\cite{RIXS1,RIXS2,RIXS3,RIXS4,RIXS5,RIXS6}. 

We will treat the coupling between the itinerant electron and local moment perturbatively. To the lowest order of the coupling the electron self-energy is given by
\begin{equation}
\Sigma(\mathrm{k},i\nu)=3\times\frac{g^{2}}{4N\beta}\sum_{\mathrm{q},i\omega}\chi(\mathrm{q},i\omega)G^{0}(\mathrm{k-q},i\nu-i\omega),\nonumber
\end{equation}
in which $G^{0}(\mathrm{k},i\nu)=(i\nu -\epsilon_{\mathrm{k}})^{-1}$ is the Green's function of the bare electron. The factor 3 comes from the three spin components. The imaginary part of the electron self-energy is given by
\begin{eqnarray}
\mathrm{Im}\Sigma(\mathrm{k},\omega)=-\frac{c}{N}\sum_{\mathrm{q}}[1+n_{B}(\omega_{\mathrm{k,q}})-f(\epsilon_{\mathrm{k-q}})]R(\mathrm{q},\omega_{\mathrm{k,q}}),\nonumber
\end{eqnarray}
in which $c=3g^{2}/16\pi$, $\omega_{\mathrm{k,q}}=\omega-\epsilon_{\mathrm{k-q}}$. $R(\mathrm{q},\omega)=-2\mathrm{Im}\chi(\mathrm{q},\omega)$ is the spectral function of the spin fluctuation. The electron spectral function as calculated from the renormalized Green's function is given by 
\begin{equation}
A(\mathrm{k},\omega)=\frac{-2\mathrm{Im}\Sigma(\mathrm{k},\omega)}{[\omega-\mathrm{Re}\Sigma(\mathrm{k},\omega)]^{2}+[\mathrm{Im}\Sigma(\mathrm{k},\omega)]^{2}}.\nonumber
\end{equation}
In our calculation, we treat $g^{2}\chi_{\mathrm{Q}}$ as a free parameter. 

The electron spectral function along the M-X line is plotted in Figure 3 for several values of the coupling constant. One find that the AF shadow band in Figure 2 is now replaced by a broad hump structure, whose spectral maximum follows more or less the same dispersion as the AF shadow band. The main band remains sharp and is pushed to higher energy. These spectral characteristics can be understood simply as a level repulsion effect between the bare quasiparticle state at $\epsilon_{\mathrm{k}}$ and the AF scattered state at $\epsilon_{\mathrm{k-q}}+\Omega_{\mathrm{q}}$. In the spin-Fermion model studied here, such a level repulsion effect is maximized at the M point, where $\epsilon_{\mathrm{k}}=\epsilon_{\mathrm{k+Q}}$ and the quasiparticle dispersion is minimized as a result of the proximity to the Van Hove singularity. This explains why quasiparticle in the anti-nodal region is the most strongly renormalized by the magnetic scattering and why the hump structure is the most intense at the M point.
\begin{figure}
\includegraphics[width=8cm]{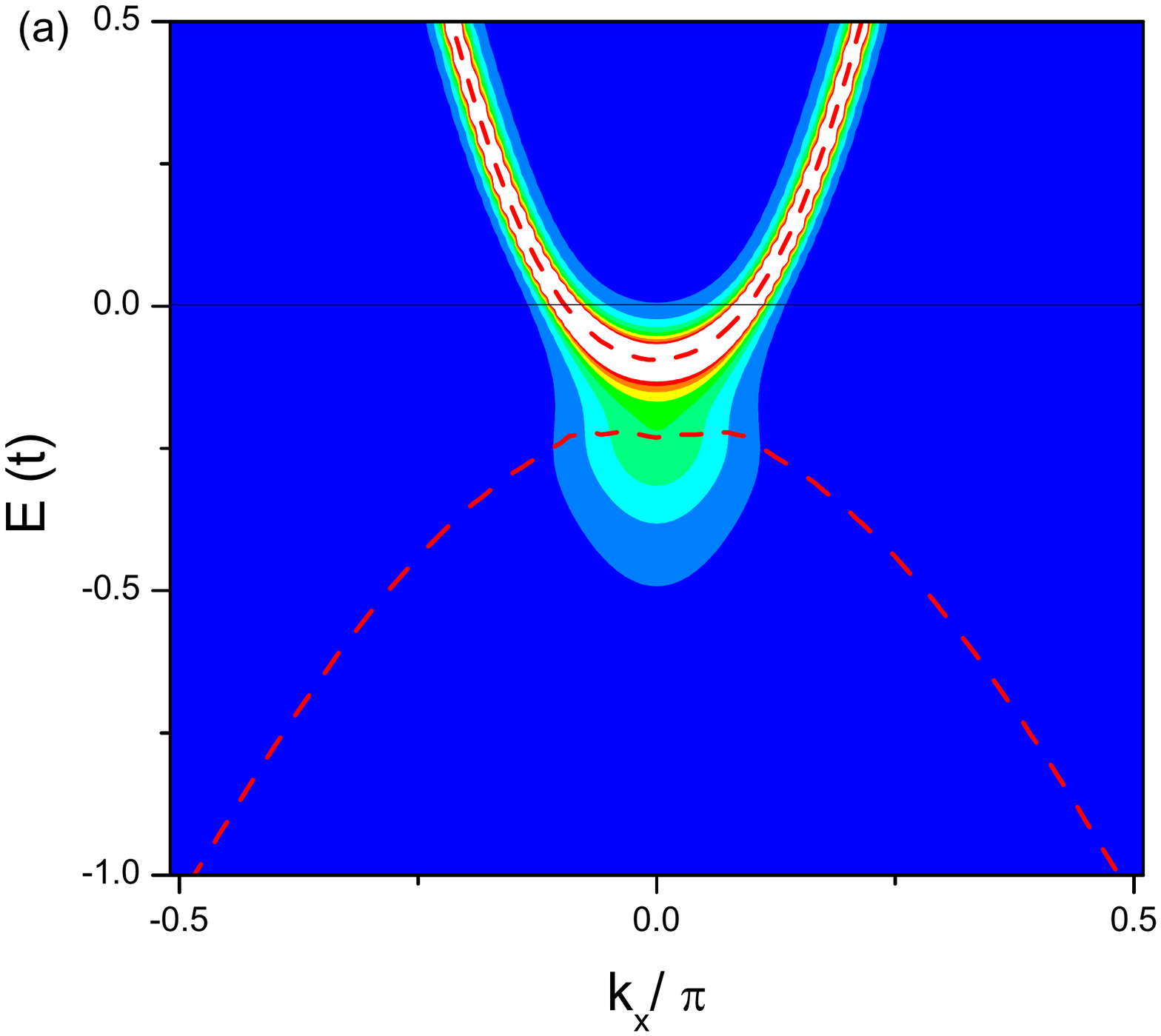}
\includegraphics[width=8cm]{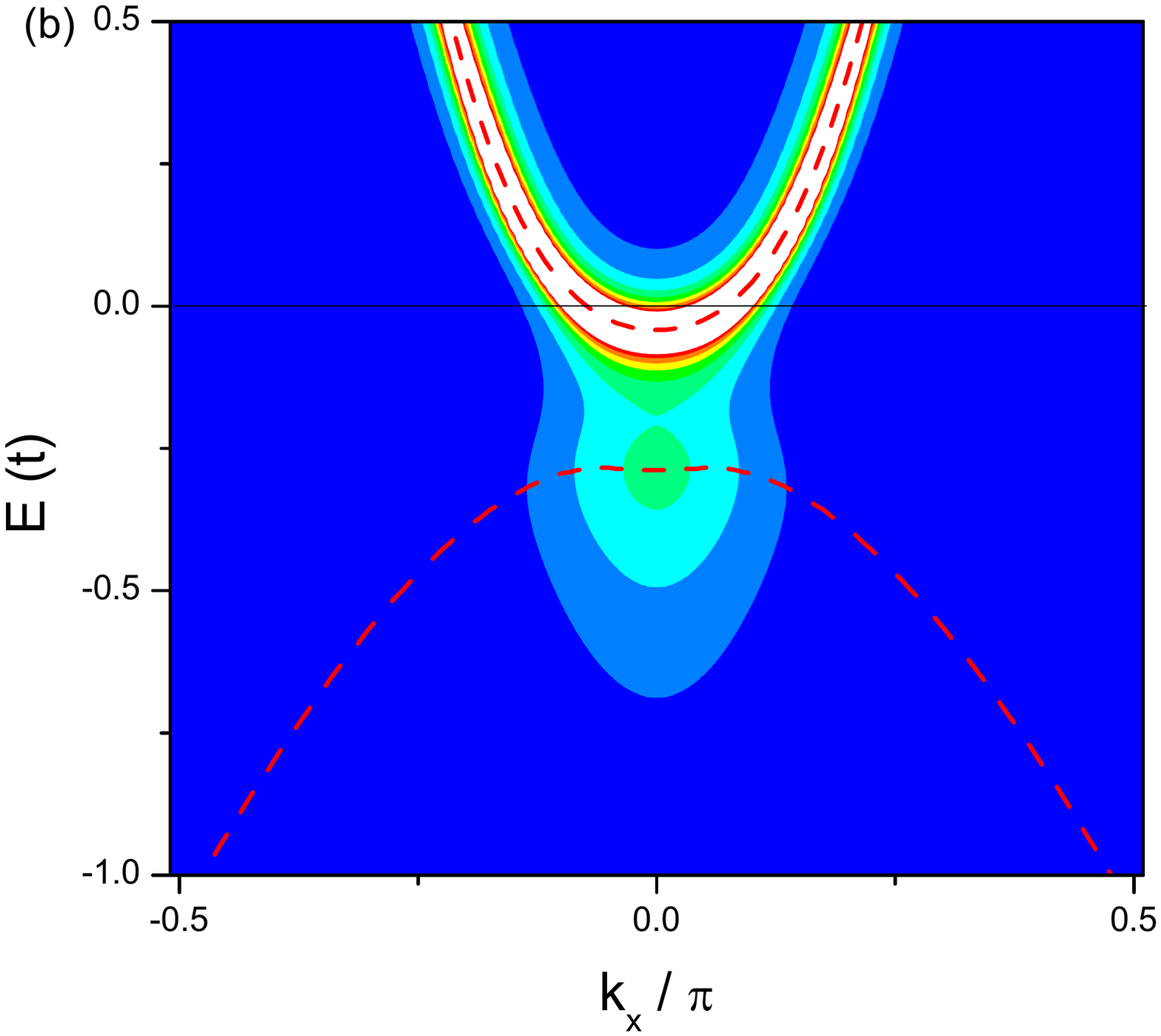}
\includegraphics[width=8cm]{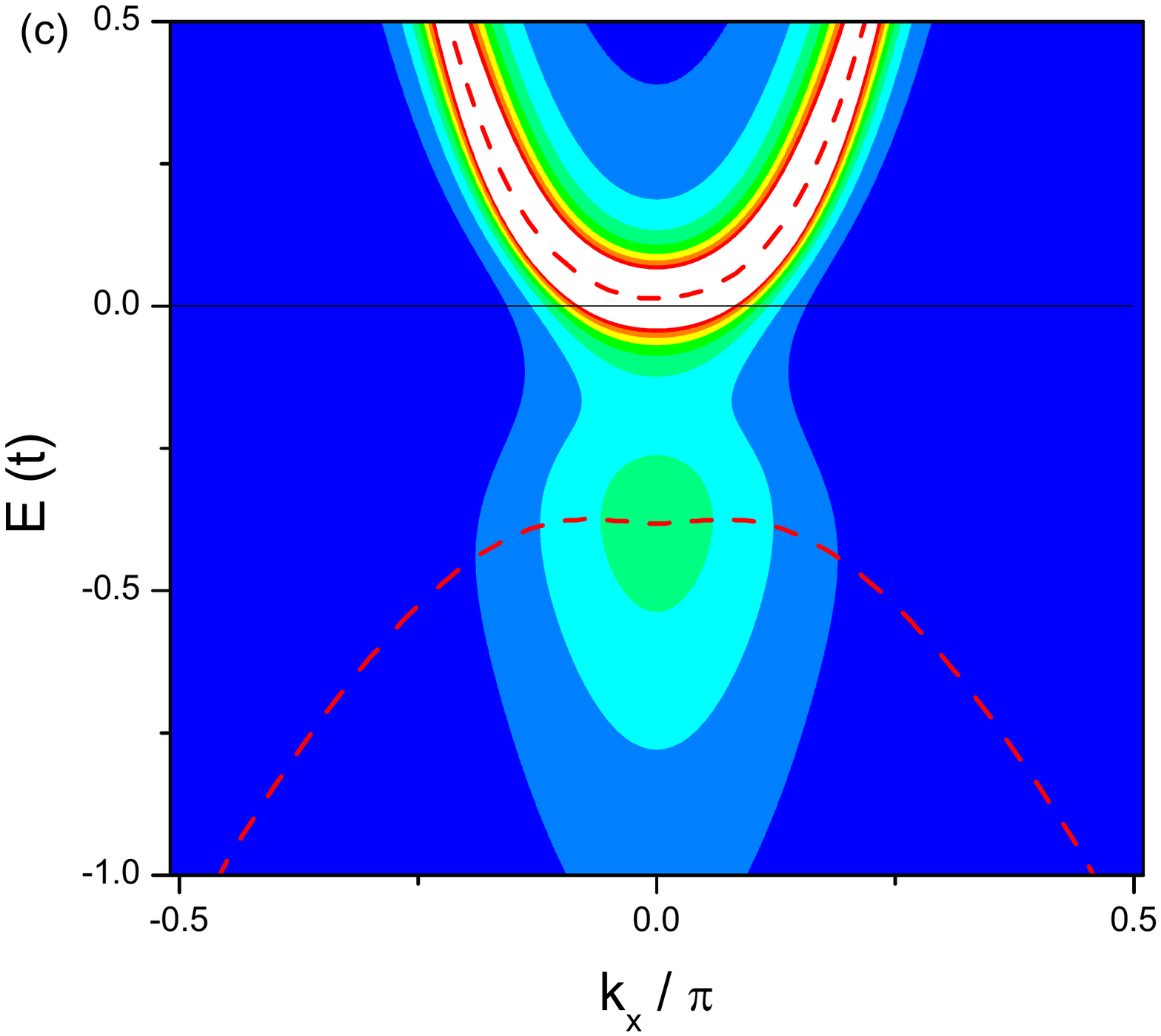}
\includegraphics[width=8cm]{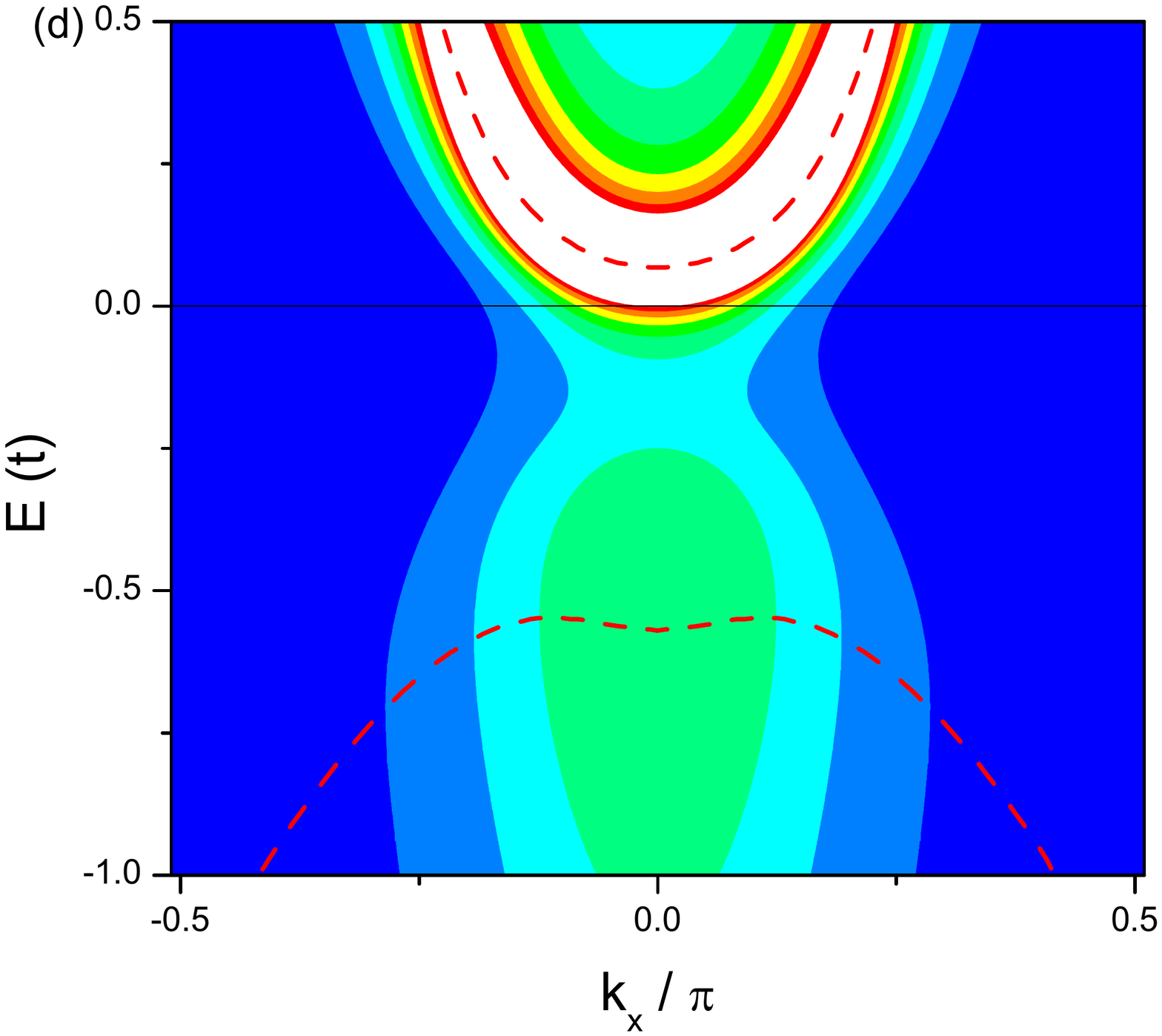}
\caption{\label{fig1}
(Color on-line) The electron spectral function along the M-X line for $3g^{2} \chi_{\mathrm{Q}}/\pi$ equal to 200, 400, 800 and 1600 in unit of $t$.\cite{coupling} The red dashed lines mark the positions of the spectral maximum. The black line marks the position of the Fermi level. }
\end{figure}

\begin{figure}
\includegraphics[width=8cm]{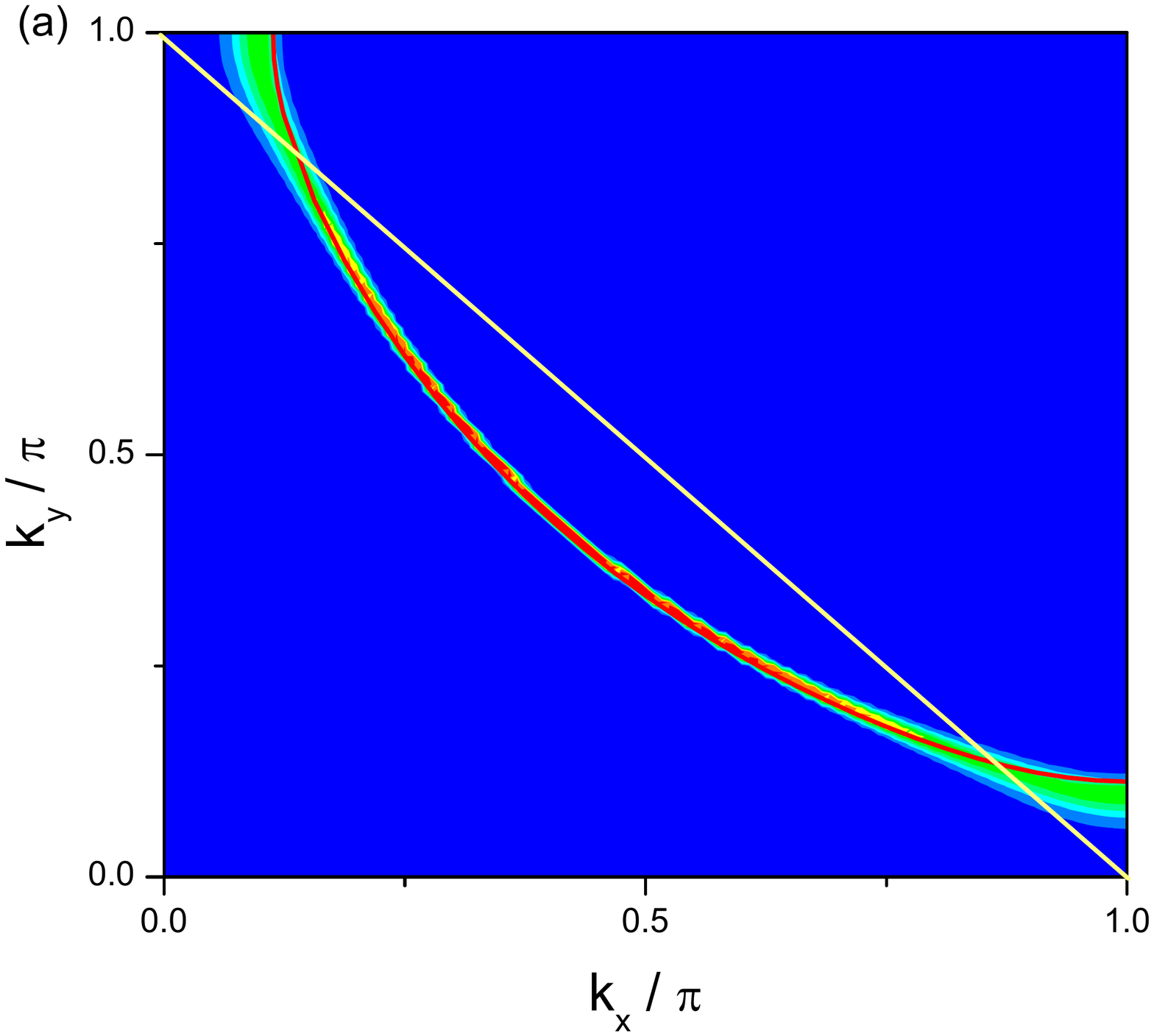}
\includegraphics[width=8cm]{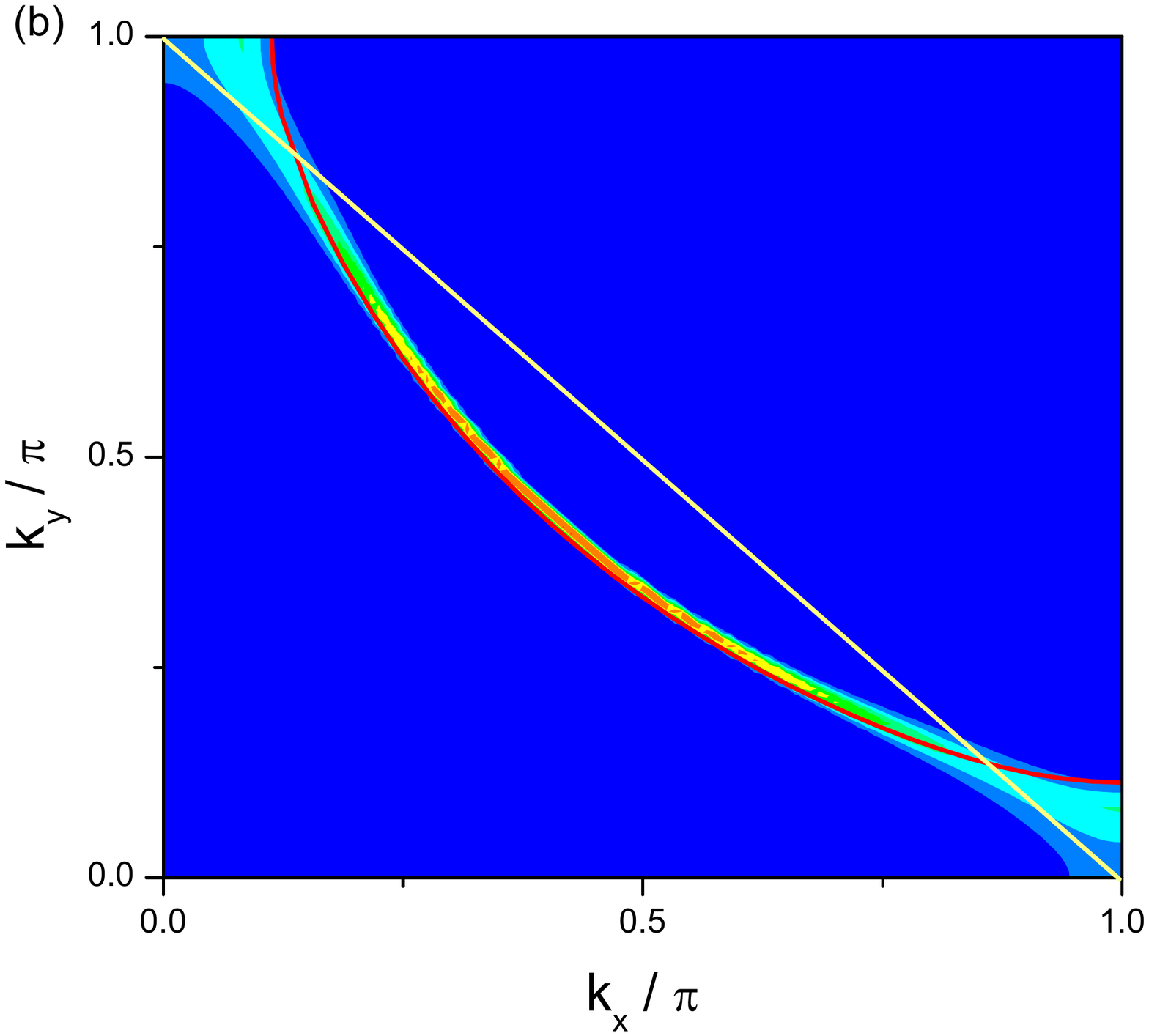}
\includegraphics[width=8cm]{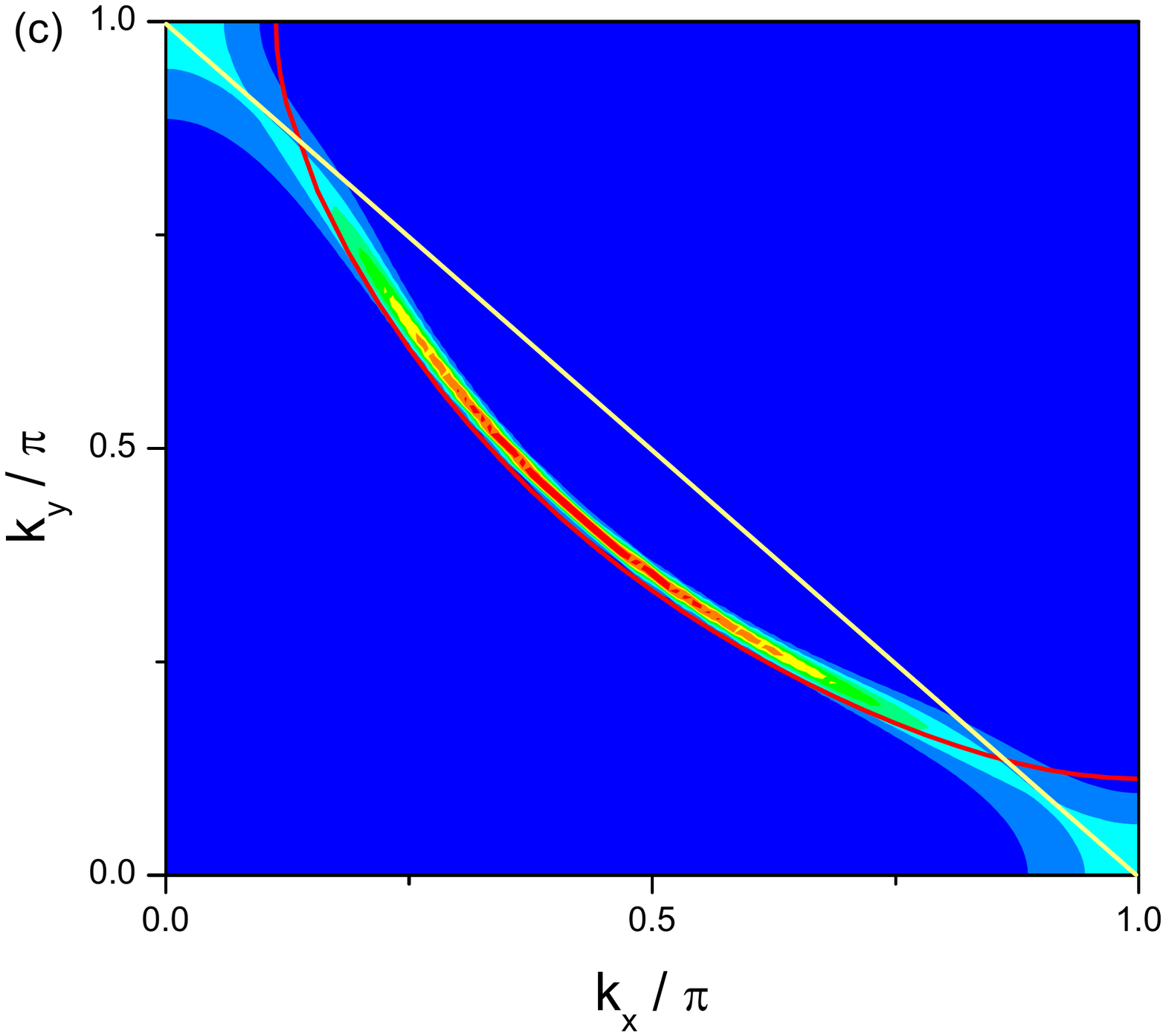}
\includegraphics[width=8cm]{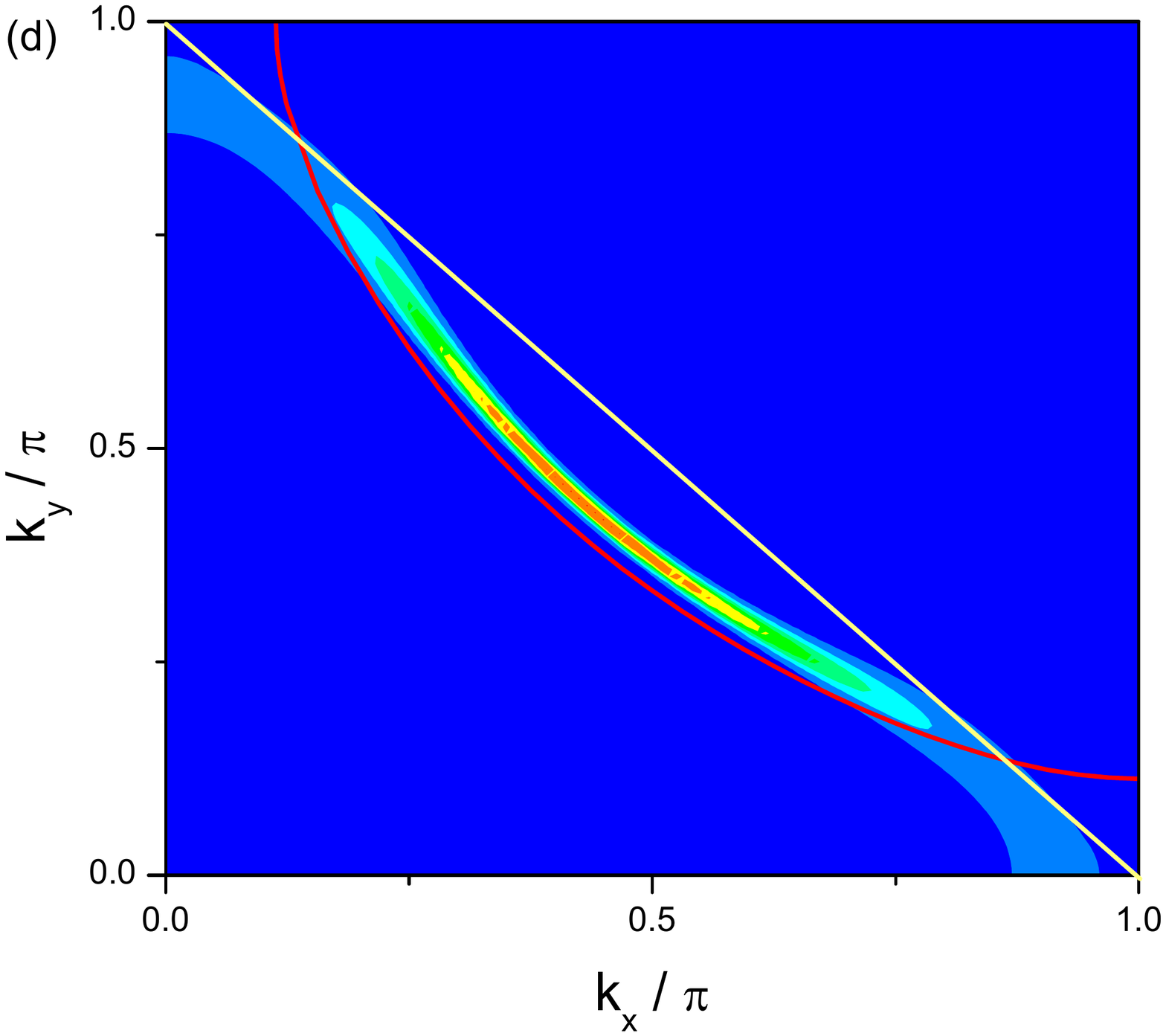}
\caption{\label{fig1}
(Color on-line) The Fermi surface map for $3g^{2} \chi_{\mathrm{Q}}/\pi$ equal to 200, 400, 800 and 1600 in unit of $t$. The red line marks the bare Fermi surface. Plotted in the figure is the integrated spectral weight in an energy window of width 20 meV around the Fermi level. The yellow line marks the boundary of the AF Brillouin zone.}
\end{figure}

According to the above picture, the Fermi level crossing along the M-X line can only be eliminated when the M point is pushed above the Fermi level. Thus, if the pseudogap is opened in a continuous way, we should expect the Fermi momentum along the M-X line to decrease to zero as the temperature is lowered toward T*. This is obviously at odds with the ARPES observation, according to which the Fermi momentum along the M-X line is almost temperature independent above T* and the M point is always in the occupied side. We note that such an inconsistency exists not only in the AF spin fluctuation scenario, but exists in all competing order scenarios with an order parameter that preserve the U(1) charge conservation. The reason is as follows. While a competing order in the particle-hole channel can renormalize the quasiparticle dispersion, it can not eliminate the distinction between an occupied and an unoccupied state. If the M and X point are on the opposite side of the Fermi level, it then follows from continuity consideration that the quasiparticle energy must cross the Fermi level somewhere along the M-X line. This is in stark contrast with the situation in the pairing scenario, when the quasiparticle is the mixture of particle-like and hole-like object and has always a positive energy. We thus conclude the leading edge gap observed in the ARPES measurement below T* must involve electron pairing. This is the most important conclusion of this work.    

Although the magnetic scattering alone can not account for the observed leading edge gap in the ARPES spectrum below T*, it is indispensable for a consistent understanding of the quasiparticle dynamics in the cuprates. In particular, the emergence of the high energy hump structure in the electron spectrum and the back bending of its spectral maximum, which occurs at a momentum different from the underlying Fermi momentum, should both be attributed to the AF band folding effect. In the following, we will look more closely on the consequence of the AF band folding effect.

In Figure 4, we map out the spectral weight on the Fermi energy for several values of $g^{2}\chi_{\mathrm{Q}}$. A single large Fermi surface is observed for all values of $g^{2}\chi_{\mathrm{Q}}$. The AF shadow band simply does not appear in the Fermi surface mapping. At the same time, the Fermi surface expands in the nodal region and shrinks in the anti-nodal region. The renormalized Fermi surface thus becomes better AF nested. These spectral characteristics can be understood as follows. First, in the background of a short-ranged and dynamical spin fluctuation, the AF shadow band is smeared out in energy as a result of the dispersion in the quasiparticle energy and the diffusion in the spin fluctuation energy. Second, since $\epsilon_{\mathrm{k}}<\epsilon_{\mathrm{k+Q}}$ for $\mathrm{k}$ within the AF Brillouin zone, the quasiparticle energy is expected to be lowered in the nodal region and lifted in the anti-nodal region by the level repulsion effect.  The Fermi surface should thus bend outwards in the nodal region and bend inwards in the anti-nodal region and thus become more susceptible to AF ordering. We thus expect the AF band folding effect to enhance the magnetic susceptibility at $\mathrm{Q}=(\pi,\pi)$.

Below T*, we should consider both the AF band folding effect and the electron pairing effect induced by the AF spin fluctuation simultaneously. In a recent work of us, we find that the AF band folding effect is responsible for the unusually flat quasiparticle dispersion in the anti-nodal region below T$_{c}$\cite{bcs-flat}. In fact, both effects can be understood as self-energy correction to the quasiparticle dynamics induced by the AF spin fluctuation and should thus be treated on an equal footing.   According to this picture, the electron spectrum of the pseudogap phase should be similar in nature to that of the superconducting phase. In particular, we expect the peak-dip-hump structure in the anti-nodal region to develop just below T*, although it becomes prominent only below T$_{c}$.

At the same time, the AF band folding effect may act as the driving force of the pseudogap phenomena. More specifically, the strong quasiparticle dressing by AF spin fluctuation will greatly reduce the kinetic energy penalty for electron pairing in the anti-nodal region, making possible the development of pairing gap at a temperature significantly higher than that in the nodal region(which is nothing but T$_{c}$). In other words, the development of the pairing gap below T* can be understood as a process through which the system reorganize its spin fluctuation spectrum so as to reduce the electron incoherence caused by the scattering from the low energy spin fluctuations and enhance the electron pairing mediated by higher energy spin fluctuations.

The scenario presented above has many similarities with the strong coupling theory of phonon-mediated superconductivity. In particular, in both cases the self-energy correction in the normal channel is playing an important role for electron pairing.  However, the two theories differ from each other on the following important points. First, the spin fluctuation in the magnetic scenario is subjected to strong feedback effect from electron pairing, while the phonon is almost free from such a feedback effect since it is an independent degree of freedom. Second, the self-energy correction induced by the AF scattering is strongly anisotropic on the Fermi surface in both the normal and the anomalous channel, while those induced by phonon scattering is essentially isotropic on the Fermi surface. Third, in the AF scattering scenario the Migdal theorem is strongly violated in the anti-nodal region as a result of the proximity to the Van Hove singularity, while in the case of phonon-mediated superconductivity the Migdal theorem is usually valid on the whole Fermi surface. A full theory of the quasiparticle dynamics in the magnetic scenario should thus include the contribution from higher order scattering process in a self-consistent way with proper account of the vertex correction effect. On a qualitative level, the higher order scattering process will push the spectral maximum of the hump structure to higher binding energy and reduce the extend of band renormalization. On the other hand, the vertex correction will enhance the coupling between the AF spin fluctuation and the quasiparticle, especially so in the anti-nodal region. We leave a more detailed analysis of the physical consequence of these differences to future studies.

In conclusion, through the study of quasiparticle dynamics under the scattering of short-ranged and dynamical spin fluctuation, we show that electron pairing is indispensable for the development of the leading edge gap observed in ARPES measurement below T*. Nevertheless, we find the AF band folding effect is important for the understanding of the quasiparticle dynamics in the pseudogap phase. The frustration of kinetic energy related to the AF band folding effect may even provide a driving force for the development of the pairing gap in the anti-nodal region, in which the quasiparticle is strongly dressed by the AF spin fluctuation as a result of proximity to the Van Hove singularity. Finally, we note that it is more illuminating to search for the evidence of particle-hole symmetry breaking in energy space, rather than in momentum space.


\begin{thebibliography}{99}

\bibitem{Timusk} T. Timusk and B. Statt, Rep. Prog. Phys. \textbf{62}, 61 (1999).
\bibitem{ARPES1}A. Damascelli, Z. Hussain and Z.X. Shen, Rev. Mod. Phys. \textbf{75}, 473(2003).
\bibitem{Lee} P. A. Lee, N. Nagaosa and X.G.Wen, Rev. Mod. Phys., \textbf{78},17 (2006).
\bibitem{Kordyuk}A. A. Kordyuk, Low Temperature Physics \textbf{41}, 5 (2015).


\bibitem{Orenstein}J. Corson, R. Mallozzi, J. Orenstein, J. N. Eckstein and I. Bozovic, Nature \textbf{398}, 221 (1999).
\bibitem{Xu}Z. A. Xu, N. P. Ong, Y. Wang, T. Kakeshita, and S. Uchida, Nature 􏰀􏰁\textbf{406}, 486 (􏰀2000).􏰁
\bibitem{Lu}Lu Li, Yayu Wang, Seiki Komiya, Shimpei Ono, Yoichi Ando, G. D. Gu, and N. P. Ong
Phys. Rev. B \textbf{81}, 054510(2010).

\bibitem{Weak}J. Schmalian, D. Pines, and B. Stojkovic, Phys. Rev. Lett. \textbf{80}, 17 (1998).


\bibitem{Particle}M. Hashimoto, R.-H. He, K. Tanaka, J.-P. Testaud, W. Meevasana, R. G. Moore, D.H. Lu, H. Yao, Y. Yoshida, H. Eisaki, T. P. Devereaux, Z. Hussain and Z.-X. Shen, Nat. Phys. \textbf{6}, 414 (2010).

\bibitem{ARPES6}R.-H. He, M. Hashimoto, H. Karapetyan, J. D. Koralek, J. P. Hinton, J. P. Testaud, V. Nathan, Y. Yoshida, H. Yao, K. Tanaka, W. Meevasana, R. G. Moore, D. H. Lu, S.-K. Mo, M. Ishikado, H. Eisaki, Z. Hussain, T. P. Devereaux, S. A. Kivelson, J. Orenstein, A. Kapitulnik, Z.-X. Shen, Science \textbf{331}, 1579 (2011).

\bibitem{ARPES7}M. Hashimoto, I. M. Vishik, R.-H. He, T. P. Devereaux and Z.-X. Shen, Nat. Phys. \textbf{10}, 483 (2014).
\bibitem{Zheng}G.-Q. Zheng, P. L. Kuhns, A. P. Reyes, B. Liang and C. T. Lin, Phys. Rev. Lett. \textbf{94}, 047006(2005).



\bibitem{RIXS1}M. Le Tacon, G. Ghiringhelli, J. Chaloupka, M. M. Sala,
V. Hinkov, M. W. Haverkort, M. Minola, M. Bakr, K. J. Zhou, S.
Blanco-Canosa, C. Monney, Y. T. Song, G. L. Sun, C. T. Lin, G. M. De
Luca, M. Salluzzo, G. Khaliullin, T. Schmitt, L. Braicovich, and B.
Keimer, Nat. Phys. \textbf{7}, 725 (2011).

\bibitem{RIXS2}M. P. M. Dean, R. S. Springell, C. Monney, K. J. Zhou,
J. Pereiro, I. Bo\v{z}ovi\'{c}, B. Dalla Piazza, H. M. R{\o}nnow, E.
Morenzoni, J. van den Brink, T. Schmitt, and J. P. Hill, Nat. Mater.
\textbf{11}, 850 (2012).

\bibitem{RIXS3}M. P. M. Dean, A. J. A. James, R. S. Springell, X. Liu,
C. Monney, K. J. Zhou, R. M. Konik, J. S. Wen, Z. J. Xu, G. D. Gu,
V. N. Strocov, T. Schmitt, and J. P. Hill, Phys. Rev. Lett.
\textbf{110}, 147001 (2013).

\bibitem{RIXS4}M. Le Tacon, M. Minola, D. C. Peets, M. Moretti Sala, S.
Blanco-Canosa, V. Hinkov, R. Liang, D. A. Bonn, W. N. Hardy, C. T.
Lin, T. Schmitt, L. Braicovich, G. Ghiringhelli, and B. Keimer,
Phys. Rev. B \textbf{88}, 020501(R) (2013).

\bibitem{RIXS5}M. P. M. Dean, G. Dellea, R. S. Springell, F. Yakhou-
Harris, K. Kummer, N. B. Brookes, X. Liu, Y.-J. Sun, J. Strle, T.
Schmitt, L. Braicovich, G. Ghiringhelli, I. Bozovic, and J. P. Hill,
Nat. Mater. \textbf{12}, 1019-2023 (2013)

\bibitem{RIXS6}M. P. M. Dean,  G. Dellea, M. Minola, S. B. Wilkins, R. M. Konik, G. D. Gu, M. Le Tacon,
N. B. Brookes, F. Yakhou-Harris, K. Kummer, J. P. Hill, L.
Braicovich, and G. Ghiringhelli, Phys. Rev. B \textbf{88}, 020403(R) (2013).

\bibitem{NAFL1}A. Abanov, A. V. Chubukov, and J. Schmalian, Advances in Physics \textbf{52}, 119(2003).
\bibitem{NAFL2}P. Monthoux, A. V. Balatsky and D. Pines, Phys. Rev. Lett. \textbf{67}, 3448 (1991).
\bibitem{NAFL3}P. Monthoux and D. Pines: Phys. Rev. Lett. \textbf{69}, 961(1992).
\bibitem{NAFL4}A. V. Chubukov, D. Pines and J. Schmalian, arXiv:0201140.
\bibitem{NAFL5}J. Schmalian, D. Pines, and B. Stojković, Phys. Rev. Lett. \textbf{80}, 3839 (1998); Phys. Rev. B \textbf{60}, 667(1999).

\bibitem{edop}Tao Li and Da-Wei Yao, arXiv:1803.08226.

\bibitem{MMP1}A. Millis. H. Monien and D. Pines, Phys. Rev. B \textbf{42}, 197 (1990).
\bibitem{MMP2}Y. Zha, V. Barzykin and D. Pines, Phys. Rev. B \textbf{54}, 2561 (1996).

\bibitem{coupling}Using the estimation of the hyperfine coupling constant listed in table III of reference [26] and the spin relaxation rate data presented in reference [12], we expect that $3g^{2}\chi_{\mathrm{Q}}/\pi=1000$ when $g=2.5t$ around T*, if we assume $\xi=3$ and $\omega_{sf}=15 \mathrm{meV}$. We note that four incommensurate peaks around Q are assumed in reference [26] and that the hyperfine coupling constant on the copper site is only weakly material dependent.  
\bibitem{bcs-flat}Tao Li and Da-Wei Yao, arXiv:1805.04883.





\

\end{thebibliography}
\end{document}